# AN IMPROVISED ALGORITHM TO IDENTIFY THE BEAUTY OF A PLANAR CURVE

Gobithaasan, R.U.
*Dept. of Mathematics, FST, University Malaysia Terengganu, 21030 Kuala Terengganu, Malaysia.*
*gr@umt.edu.my or gobithaasan@gmail.com*

Jamaludin Md. Ali
*School of Mathematical Sciences, Universiti Sains Malaysia, 11800 Minden, Penang, Malaysia.*
*jamaluma@cs.usm.my*

Kenjiro T. Miura
*Dept. of Mechanical Engineering, Shizuoka University, 3-5-1, Johoku, Hamamatsu,Shizuoka, Japan.*
*tmkmiur@ipc.shizuoka.ac.jp*

**Abstract :**

*An improvised algorithm is proposed based on the work of Yoshimoto & Harada (2002). The improvised algorithm results a graph which is called LDGC or Logarithmic Distribution Graph of Curvature. This graph has the capability to identify the beauty of monotonic planar curves with less effort as compared to LDDC by Yoshimoto & Harada (2002).*

## Introduction

A potential customer judges the aesthetic appeal of a product before the physical performance. This clearly indicates the importance of aesthetic shapes for the success of an industrial product. In curve design environment, a curve is characterized based on its curvature profile. A curvature profile is a graph plotted with the values of parameter $t$ representing x-axis against its corresponding signed curvature values representing y-axis and it is used widely as a shape interrogation tool to fair B-spline curves and surfaces. The designer arrives to the desired curve by interactively or automatically tweaking the control points and concurrently inspecting the curvature profile.

There are many studies indicate the importance of curvature profile to characterize planar curves (see Nutbourne & Martin (1980) and references therein). Curvature profile has been highlighted as the shape interrogation tool to fair B-spline curves and surfaces in (Farin & Sapidis, 1989). Since then, it has been the de facto standard to verify the fairness of a curve. By interactive or automated tweaking control points and concurrently inspecting the curvature profile, the designer arrives to the desired curve. However, curvature cannot be used to identify the beauty of a planar curve.

A different kind of approach has been proposed by Yoshimoto & Harada (2002) to analyze the characteristics of planar curves with monotonic curvature. The relationship between the length frequencies of a segmented curve with regards to its radius of curvature is plotted in log-log coordinate system and named as Logarithmic Distribution Diagram of Curvature or LDDC. It is said that these types of graphs can be used to identify the aesthetic value of a curve (see Yoshimoto & Harada (2002) for details). Harada et. al (1999) first used LDDC as a tool to characterize the curves used for automobile design. To note, the generation of LDDC is through quantitative method.

The notion behind generating LDDC is to mathematically obtain the locus of the interval of radius of curvature and its corresponding length frequency. Thus, two curves with different length would generate distinct LDDC regardless of the similarities of the shape of curvature profile. For example, two circular arcs with the same radius but different length would generate similar curvature profile; nevertheless LDDC would generate different shapes (Harada et. al, 1999).

In this paper, we propose LDGC (Logarithmic Distribution Graph of Curvature), which is an enhanced version of LDDC. There are two main advantages of LDGC; firstly the division of parameter $t$ in generating LDGC instead of segmenting the curve and calculating each arc lengths. Secondly, a general formula for segregating the radius of curvature is proposed in which a user does not need prior experience of the type of curves being analyzed. These features make LDGC more effective in terms of computational power and time consumption.

**An Improvised Algorithm**

*Logarithmic Distribution Graph of Curvature*

In this section, the generation of LDDC is enhanced and simplified to obtain Logarithmic Distribution Graph of Curvature or LDGC. The graph will be drawn directly using the midpoints of the radius of curvature classes instead of using the intervals to generate a histogram.

Let a curve be defined as $C(t) = \{x(t), y(t)\}$ where $\alpha \leq t \leq \beta$. The curve is divided into $N$ number segments for LDGC generation. In Yoshimoto & Harada (2002), the first step to generate LDDC is the division of curve into equal length of $N$ number of segments: $S_i = (S_1, S_2, \ldots, S_N)$. The constitutional points corresponding to each segments are defined as $a_i = (a_1, a_2, \ldots, a_{N+1})$ where $i \in (1, 2, \ldots, N)$. The length of each curve segment is $\frac{\int_\alpha^\beta \|C'(t)\| dt}{N}$. Thus, in order to obtain the corresponding parameter value $t_i$, one would need to solve: $\int_{t_{i-1}}^{t_i} \|C'(t)\| dt = \frac{\int_\alpha^\beta \|C'(t)\| dt}{N}$ for $i \in (1, 2, \ldots, N)$ which is clearly troublesome. This step is modified in generating LDGC as follows:

**[LDGC-1]**: Since first step of LDDC is computationally expensive, an alternative is to divide parameter $t_i$, into $N$ number of intervals:

$$t_i = \alpha; \; i = 1,$$
$$T_{N+1} = \beta; \; i = N+1, \quad (1)$$
$$t_i = t_{i-1} + \frac{\|\beta - \alpha\|}{N}; \; for \; i = \{2, 3, \ldots, N-1\}$$

When $N \to 10,000$, the length of the segments becomes equivalent, whereby equal length of curve divisions are generated.

**[LDGC-2]** Calculate the average of the radius of curvature for each segment: $\bar{\rho}_i = \frac{\rho(t_i) + \rho(t_{i+1})}{2}$. Let $\tilde{\rho}_i = \log\left(\frac{\bar{\rho}_i}{\int_\alpha^\beta \|C'(t)\| dt}\right)$.

**[LDGC-3]** Now, $\tilde{\rho}_i$ will be grouped in 100 number of classes of intervals:

$$class_i = \{Min[\tilde{\rho}] + \Delta_c(i-1), Min[\tilde{\rho}] + (\Delta_c(i))\}; for \; i = \{1, 2, \ldots, N\} \quad (2)$$

where $\Delta_c = \frac{[Max[\tilde{\rho}] - Min[\tilde{\rho}]]}{100}$, with $Min[\tilde{\rho}]$ and $Max[\tilde{\rho}]$ as the minimum and maximum value occurs in $\tilde{\rho}_i$ respectively. Let the midpoint values of $class_i$ be defined as $\tilde{\rho}_i$.

**[LDGC-4]** Segregate and calculate the number of occurrence of $\tilde{\rho}_i$ according to the classes obtained from step **[LDGC-3]** and denote it as $\tilde{N}_j$, where $j \in (1, 2, \ldots, 100)$. Thus, $\sum_{j=1}^{100} \tilde{N}_j = N$.

**[LDGC-5]** Calculate the length of frequency of each class using:

$$s_j = \tilde{N}_j \cdot \frac{\int_\alpha^\beta \|C'(t)\| dt}{N} \quad (3)$$

Hence, $\sum_{j=1}^{100} s_j = \int_\alpha^\beta \|C'(t)\| dt$.

**[LDGC-6]** Next, calculate:

$$\tilde{s}_j = \log \left( \frac{s_j}{\int_a^b \|c'(t)\| dt} \right) \qquad (4)$$

**[LDGC-7]** Finally, LDGC can be generated by representing $\tilde{\rho}_i$ for horizontal axis and its' corresponding values of $\tilde{s}_j$ for vertical axis.

*Identification of Beautiful Curves*

The gradient of LDGC is defined as follows:

$$gradient = \frac{ds_j}{d\tilde{\rho}_i} = \lim (s_{j-1} - s_j)/(\tilde{\rho}_{i-1} - \tilde{\rho}_i) \qquad (5)$$

If the gradient is equivalent to a constant value, then the curve is said to have self affine property. The beauty of a curve increases as the gradient of the curve approximates to a constant value. Refer to Yoshimoto & Harada (2002) for the grouping of aesthetic curves.

**Numerical Examples**

*Clothoid*

The clothoid is defined in plane as parametric equation in terms of Fresnel integrals by:

$$\begin{pmatrix} x(t) \\ y(t) \end{pmatrix} = \pi B \begin{pmatrix} FC(t) \\ FS(t) \end{pmatrix} \qquad (6)$$

where the scaling factor $\pi B$ is positive, parameter t is nonnegative and the Fresnel integrals are defined as:

$$FC(t) = \int_0^t \cos \frac{\pi u^2}{2} \qquad (7)$$

$$FS(t) = \int_0^t \sin \frac{\pi u^2}{2} \qquad (8)$$

Since the radius of curvature of clothoid is given by $\rho = B/t$, the linear curvature profile will generate a straight line of LDGC. Figure 1 illustrates the clothoid curve when $t = 1.5$ and $b = 1$.

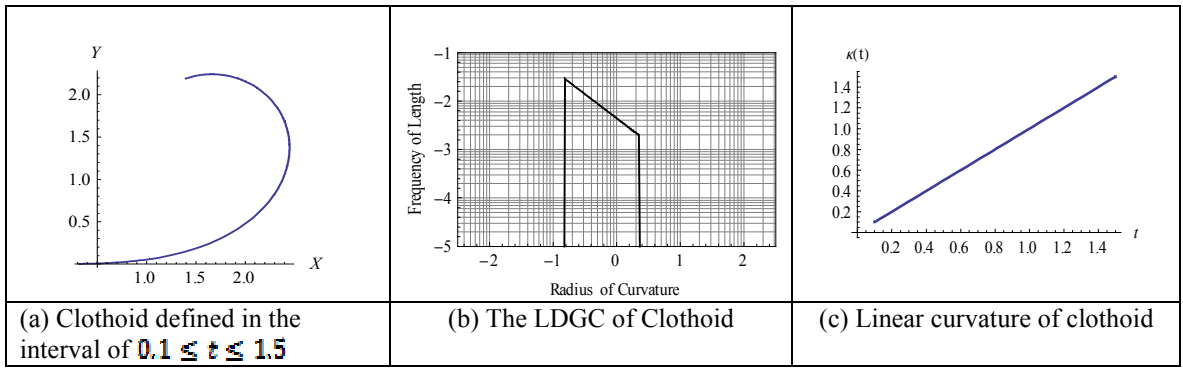

| (a) Clothoid defined in the interval of $0.1 \leq t \leq 1.5$ | (b) The LDGC of Clothoid | (c) Linear curvature of clothoid |

**Figure 1** : Clothoid, its LDGC and curvature profile.

*Circle Involute*

A circle involute is defined in a plane as:

$$C(t)=Cos(t)+t\,Sin(t),\ Sin(t)-t\,Cos(t) \qquad (9)$$

where parameter *t* represents the winding angle of a circle. Figure 2 illustrates the curve, its LDGC and curvature profile.

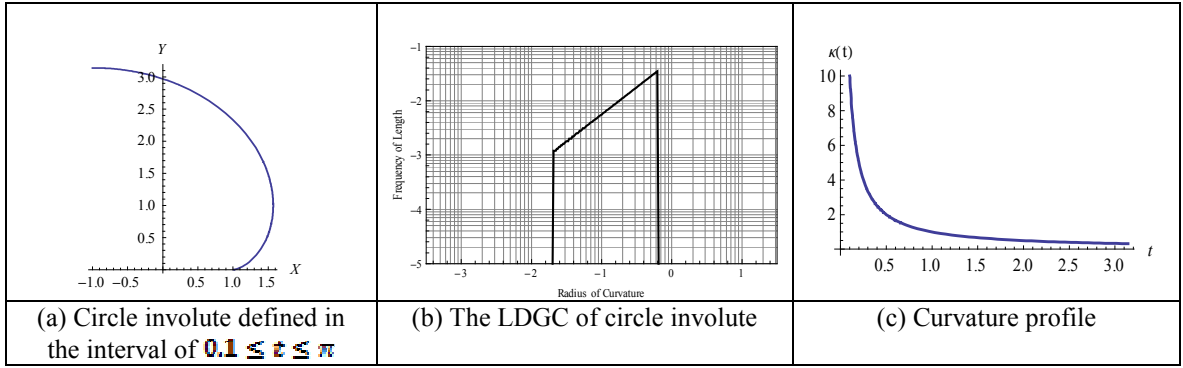

| (a) Circle involute defined in the interval of $0.1 \leq t \leq \pi$ | (b) The LDGC of circle involute | (c) Curvature profile |

**Figure 2** : Circle involute, its LDGC and curvature profile.

*Bezier Cubic*

A Bezier cubic is defined in plane as:

$$B(t) = (1-t)^3 A_0 + 3t(1-t)^2 A_1 + 3t^2(1-t) A_2 + t^3 A_3 \qquad (10)$$

where $0 \leq t \leq 1$ and $A_i = \{x_i, y_i\}$ are its control points. In Figure 3, the LDGC for Bezier cubic with the following control points is generated: $A_0 = \{0,1\}, A_1 = \{1,1\}, A_2 = \{2,1\}$ and $A_3 = \{3,1.5\}$.

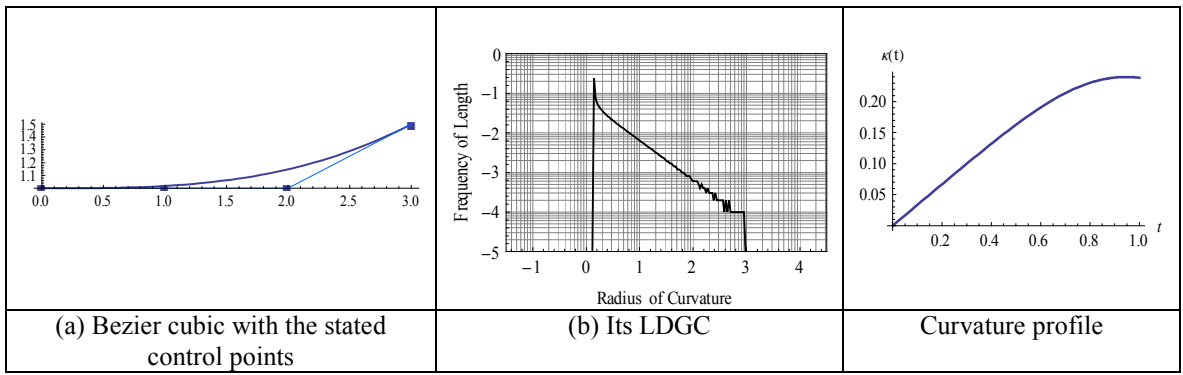

| (a) Bezier cubic with the stated control points | (b) Its LDGC | Curvature profile |

**Figure 1** : Bezier cubic, its LDGC and curvature profile.

## Conclusion

There are two enhancement of LDDC in which it is now denoted as LDGC. Firstly, the proposed algorithm by Yoshimoto & Harada (2002) is computationally expensive in particular, if a curve is segmented into *N* number of segments, we need to solve *N* number of solutions in order to obtain the corresponding constitutional points. To note, a curve must be segmented into 10,000 or higher number of segments to achieve an accurate LDGC (see step **[LDGC-1]** for details). As an alternative, the generation of LDGC involves the division of parameter *t*, which results a similar graph in shorter period of time; step **[LDGC-2]**.

Secondly, the segregation of radius of curvature based on the formulated classes is troublesome in which, the formulation of classes for segregation is based on the author's experience from investigating the range of curves adopted from actual automobiles (see Yoshimoto & Harada (2002) for details). Hence, a more general classification has been proposed for LDGC generation; see step **[LDGC-4].**

**Acknowledgements**

The first author acknowledges Ministry of Higher Education Malaysia and University Malaysia Terengganu for sponsoring his Ph. D studies. The authors extend their gratitude to Ministry of Science, Technology & Inovation Malaysia (MOSTI) and Universiti Sains Malaysia for providing research grant (FRGS grant No: 203/PMATHS/671192) and Mathematica which was utilized for this research.